\documentclass[ twocolumn, tighten, twocolappendix]{aastex7}

\pdfoutput=1 
\usepackage{amsmath,amssymb,amstext}
\usepackage[T1]{fontenc}
\usepackage{apjfonts}
\usepackage{ae,aecompl}
\usepackage[utf8]{inputenc}
\usepackage{hyperref}
\usepackage[figure,figure*]{hypcap}
\usepackage{url}
\usepackage{mdwlist}
\usepackage{multirow}
\usepackage{lineno}
\usepackage{fontawesome}
\usepackage{xcolor}
\usepackage{subcaption}

\usepackage{verbatim}
\usepackage{booktabs}
\usepackage{multirow}
\usepackage{rotating}

\begin{document}

\title{Mapping Atmospheric Features of the Planetary-mass Brown Dwarf SIMP 0136 with JWST NIRISS}

 \author[]{Roman Akhmetshyn}
\affiliation{Department of Physics, McGill University, Montreal, Canada; \href{mailto:roman.akhmetshyn@mail.mcgill.ca}{roman.akhmetshyn@mail.mcgill.ca}}
\email{roman.akhmetshyn@mail.mcgill.ca}  

\author[]{Étienne Artigau} 
\affiliation{Trottier Institute for Research on Exoplanets, D\'epartment de Physique, Universit\'e de Montr\'eal, Montreal, Canada}
\email{etienne.artigau@umontreal.ca}

\author[]{Nicolas B. Cowan}
\affiliation{Department of Physics, McGill University, Montreal, Canada}
\affiliation{Department of Earth \& Planetary Sciences, McGill University, Montreal, Canada}
\email{nicolas.cowan@mcgill.ca}

\author{Michael K. Plummer}
\affiliation{Department of Physics and Meteorology, United States Air Force Academy, 2354 Fairchild Drive, CO 80840, USA}
\email{michaelkplummer@gmail.com}

\author[]{Fei Wang}
\affiliation{Department of Physics, Astronomy and Mathematics, University of Hertfordshire, Hatfield, UK}
\email{f.wang5@herts.ac.uk}

\author[]{Ben Burningham}
\affiliation{Department of Physics, Astronomy and Mathematics, University of Hertfordshire, Hatfield, UK}
\email{b.burningham@herts.ac.uk}

\author[]{Bj\"orn Benneke}
\affiliation{D\'epartment de Physique, Université de Montréal, Montr\'eal, Canada}
\email{bjorn.benneke@umontreal.ca}

\author[]{Ren\'e Doyon}
\affiliation{D\'epartment de Physique, Université de Montréal, Montr\'eal, Canada}
\email{rene.doyon@umontreal.ca}

\author[]{Ray Jayawardhana}
\affiliation{Department of Physics \& Astronomy, Johns Hopkins University, Baltimore, MD, 21218, USA}
\email{jayminray@gmail.com}

\author[]{David Lafreni\`ere}
\affiliation{D\'epartment de Physique, Université de Montréal, Montr\'eal, Canada}
\email{david.lafreniere@umontreal.ca}

\author[]{Stanimir A. Metchev} 
\affiliation{Western University, Department of Physics and Astronomy, London, Ontario, Canada}
\affiliation{Western University, Institute for Earth and Space Exploration, London, Ontario, Canada}
\email{smetchev@uwo.ca}

\author[]{Jason F. Rowe}
\affiliation{Department of Physics \& Astronomy, Bishop’s University, 2600 Rue College, Sherbrooke, QC J1M 1Z7, Canada}
\email{jrowe@ubishops.ca}

\begin{abstract}

In this paper, we analyze James Webb Space Telescope Near-Infrared Imager and Slitless Spectrograph time-series spectroscopy data to characterize the atmosphere of the planetary-mass brown dwarf \hbox{SIMP J01365662+093347}. Principal component analysis reveals that 81\% of spectral variations can be described by two components, implying that variability within a single rotational phase is induced by at least three distinct spectral regions. By comparing our data to a grid of Sonora Diamondback atmospheric models, we confirm that the time-averaged spectrum cannot be explained by a single model but requires a linear combination of at least three regions. Projecting these models onto the principal component plane shows that the overall variability is highly correlated with changes in temperature, cloud coverage, and possibly effective metallicity. We also extract brightness maps from the lightcurve and establish north--south asymmetry in the atmosphere. A combined multidimensional analysis of spectrophotometric variability links the three spectral regions to three atmospheric layers. Forsterite cloud and water abundance at each level form unique harmonics of atmospheric variability observed in different spectral bands. Atmospheric retrievals on the time-averaged spectrum are consistent with an optically thick iron cloud deck beneath a patchy forsterite cloud layer and with the overall adiabatic curve. We also demonstrate two new analysis methods: a regionally resolved spectra retrieval that relies on multiwavelength spherical harmonics maps, and a method to constrain brightness maps using Doppler information present in the spectra. Future observations of variable brown dwarfs of higher spectral resolution or those spanning multiple rotations should help break the mapping degeneracy.

\end{abstract}


\section{Introduction}
\label{sec:intro_man}

Rotational variability is common among brown dwarfs \citep{Artigau2009, radigan12, Buenzli_2014, Radigan_2014, Metchev_2015, Lew16, artigau18}, planetary mass objects \citep{Biller_2015, Biller_2018}, and directly imaged companions \citep{Zhou_2016, Naud2017, Zhou_2020a, Zhou_2020b}, especially those of young age and low surface gravity \citep{Vos19, Vos_2023}. Changes in lightcurve morphology are commonly attributed to the dynamic formation and dissipation of clouds, radiative localized hot spots, intense vertical transport due to chemical instability, or the differential rotation of clouds at different latitudes \citep{Bailer-Jones2001}. Specific morphology during a single rotation may be due to cloud features rotating in and out of view. Outstanding questions include whether the atmospheric structure is dominated by banding, spots, or a mixture of both, and what the main driving mechanism is that causes these structures to vary in time. One way to constrain that is to measure the timescales of atmospheric variability.

\begin{figure*}
    \centering
    \includegraphics[width=1.0\linewidth]{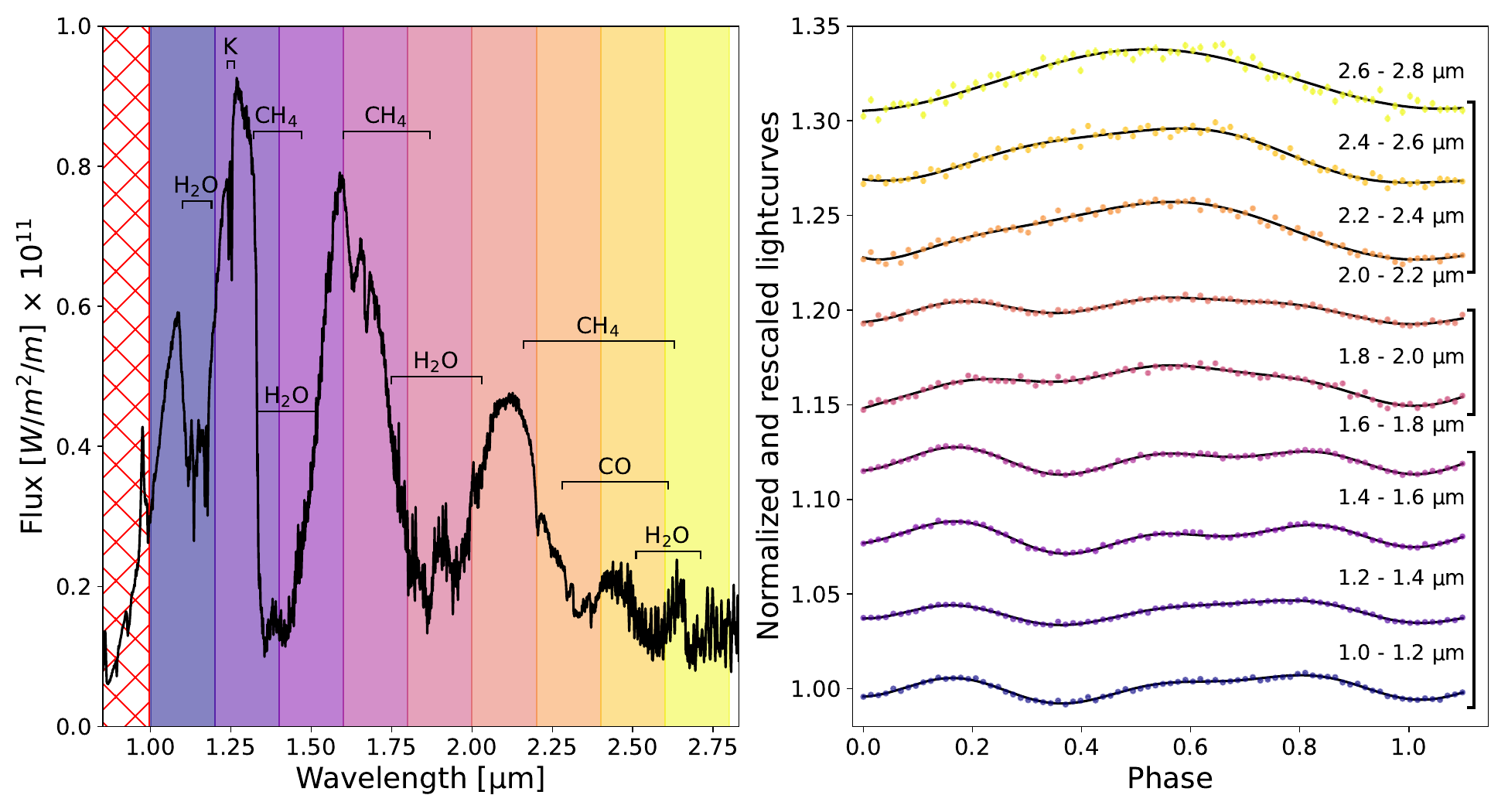}
    \caption{ Left: time-averaged spectrum of SIMP 0136 over an entire rotation with NIRISS/SOSS. The hashed red region was not considered in our analysis due to contamination of the SOSS trace from field stars. Colored bands highlight 0.2\,$\mu$m spectral bins. Right: lightcurve of each bin with best-fit \texttt{Imber} models \citep{Plummer2023,Plummer2024a}. The bins from 2.2 to 2.8 $\mu$m  (yellow) are best fit by a single peak per rotation. The intermediate bins from 1.8 to 2.2 $\mu$m  (magenta) exhibit both two and three peaks per rotation. The shortest wavelength bins from 1.0 to 1.8 $\mu$m (purple) are best fit by the third harmonic model. Fourier fitting (black lines) was performed with \texttt{Imber}.}
    \label{fig:niriss data}
    \vspace{10pt}
\end{figure*}

SIMP\,J01365662+0933473 (SIMP 0136 hereafter) is a nearby ($6.118 \pm 0.017$\,pc; \citealt{eDR3}) T2.5 brown dwarf discovered in 2006 at the Observatoire du Mont-M\'egantic with the wide-field near-infrared camera CPAPIR as part of the Sondage Infrarouge de Mouvement Propre (SIMP) proper-motion survey \citep{Artigau2006}. Further observations revealed a rotation period of $2.414 \pm 0.078$\,hr, a peak-to-peak \hbox{$J$-band} amplitude of $\sim$50\,mmag, and significant evolution of the lightcurve between each observations \citep{Artigau2009,Croll2016,Yang2016,Plummer2024b}. These properties make SIMP 0136 the brightest isolated T dwarf and an ideal target to study brown dwarf atmospheric variability. Measurements of $v \sin i$ constrained its inclination to \hbox{$80^{+10}_{-12}$}$^\circ$ \citep{Vos2017}. Further kinematic measurements identified SIMP 0136 to be part of the $\sim$200\,Myr Carina-Near moving group, which allowed estimation of its mass of $12.7 \pm 1.0$\,M$_{J}$ using brown dwarf evolutionary models \citep{Gagne2017}. The same work also estimated that SIMP 0136 could have already depleted half of its deuterium content. Given its mass at the planet--star boundary, SIMP 0136 has been called a free-floating planet. Multiple ground- and space-based observations confirmed its variability across multiple rotations \citep{Artigau2009,Croll2016,Yang2016,Plummer2024b,McCarthy2025}. General circulation models (GCMs) predict that this variability is mainly caused by patchy cloud layers \citep{Tan2021b}. It is believed that different cloud decks form and dissipate at different timescales, rotate in and out of view, and hover over each other in different configurations \citep{Apai2013,Metchev_2015,Apai2017,Vos_2023}.

Section \ref{sec:data} describes the data and its reduction. In Section~\ref{sec:pca}, we show the principal component analysis (PCA) of our data. In Section~\ref{sec:diamondback}, we compare NIRISS time-series spectra to atmospheric models. In Section~\ref{sec:retrievals}, we describe our atmospheric retrieval and compare with the results from previous works. Section~\ref{sec:mapping} describes different mapping approaches, focusing on the results from spherical harmonic mapping. Section~\ref{sec:altitude-var} explains our multidimensional study of spectral variability and connects lightcurve morphology with meteorological structure. Section~\ref{sec:reg_spec} explains how spherical harmonics maps can be used to perform extraction and retrieval of regional spectra. Section~\ref{sec:doppler} proposes to constrain brightness maps with Doppler measurements using James Webb Space Telescope (JWST) spectroscopy.

\section{Time variable data}
\label{sec:data}

The data for this research were obtained with the Near-Infrared Imager and Slitless Spectrograph in the Single Object Slitless Spectroscopy (NIRISS/SOSS) mode on board JWST as a part of the Cycle~1 Guaranteed Time Observations program \#1209 (PI: Artigau). Observations were carried out from UT 05:08:28 to 09:05:07 on 2023 July 22, 37 hr before JWST Cycle~2 program GO \#3548 (PI: Vos) that observed SIMP 0136 in NIRSpec/BOTS \citep{McCarthy2025}. We used the SOSS-Inspired Spectroscopic Extraction (SOSSISSE) pipeline for data reduction, which was previously used and described in \citet{Lim2023}.

The reduced data consist of 81 $R\sim1200$ spectra from 0.85 to 2.83\,$\mu$m taken over slightly more than one rotational period (Figure~\ref{fig:niriss data}). The spectrum changes during the object's rotation, and this variability is different for each wavelength bin associated with certain molecules, suggesting different meteorological structure at different depths.

\section{Principal component analysis of time-varying spectra}
\label{sec:pca}

\begin{figure}
    \centering
    \includegraphics[width=0.8\linewidth]{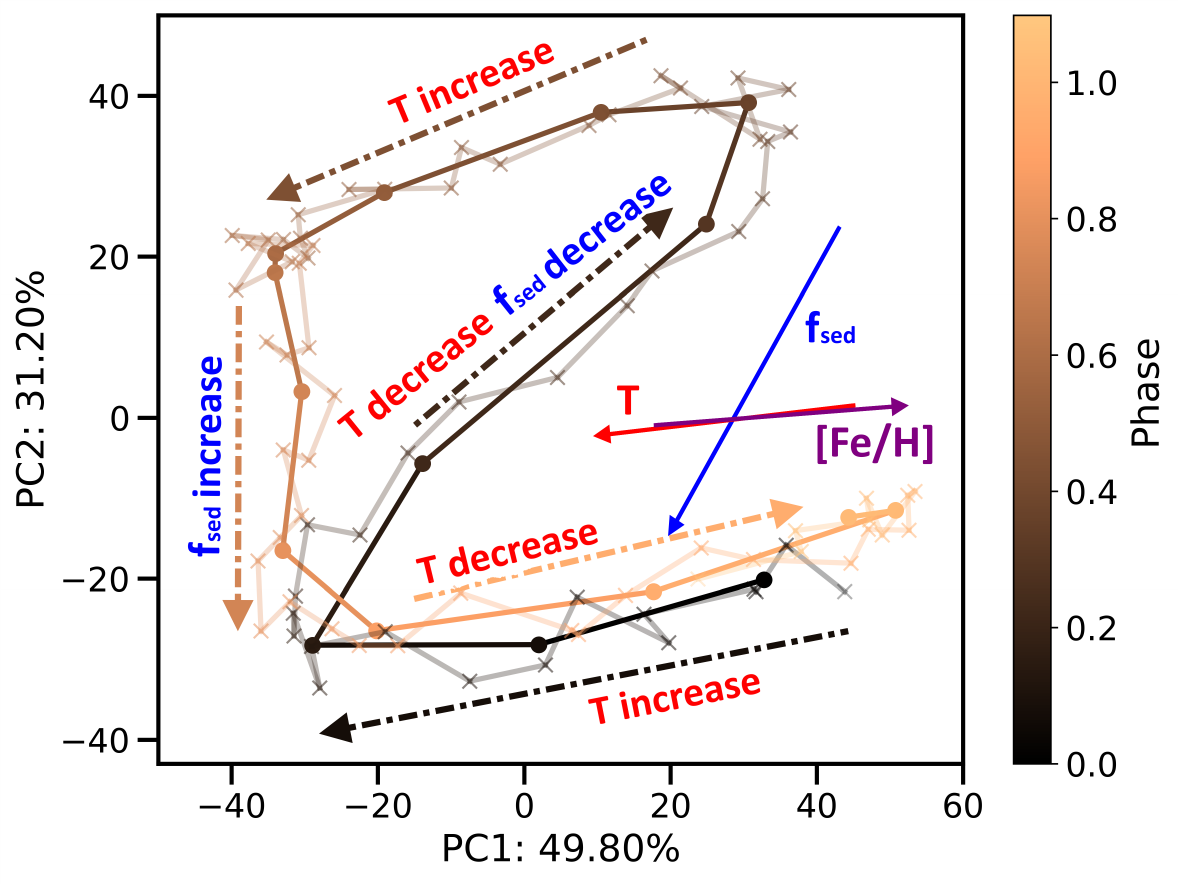}
    \caption{Projection of the time-varying spectrum onto the PCP. Percentage indicates the fraction of explained variance by the components. Low-contrast and high-contrast shaded curves, respectively, show the time series at the original time sampling and in 9 minute bins. Color-coding of both curves gives the rotational phase. The two components are orthogonal but do not directly correspond to physical parameters. We projected Sonora Diamondback models to see how changing $f_{\rm sed}$, \textit{T}, and effective [Fe/H] impacts the spectra. Starting at the bottom right of the plot, we can infer sequential change in \textit{T} and $f_{\rm sed}$ as SIMP 0136 completes a rotation.}
    \label{fig:pcp}
\end{figure}

We use principal component analysis (PCA)
to find orthogonal axes that capture the most variance in the spectrum of SIMP 0136. Since we have 81 spectra, 81 principal components can perfectly replicate all the information in the data.
We found that binning data in time does not increase the fraction of variance captured by the first two to three principal components. On the other hand, binning in wavelength greatly increases the fraction of variance explained by the first two principal components. Unbinned data have a lot of uncorrelated noise distributed over many eigenvectors, presumably Poisson noise. Averaging it out helps to focus on astrophysical variability. In the rest of this section, we use 0.019\,$\mu$m binning, which reduces the number of wavelengths to 93, while the number of principal components remains at 81.

About 81\% of the spectral variability can be explained by just two principal components. This justifies projecting our data onto the principal component plane (PCP) made of the first two principal components. According to \citet{cowan2011rotational}, the number of dimensions explaining brightness variability determines the minimum number of distinct spectral regions contributing to the variations: $N_{\text{spec}} \geq N_{\text{dim}} + 1$. These spectral regions could represent different cloud decks or clear patches.

The PCP is a useful tool for interpreting the time evolution of spectra \citep{simplex, cowan2013light}. In Figure~\ref{fig:pcp}, each datum point represents a full spectrum at a given time. From \citet{cowan2013determining}, if different spectral regions combine linearly to determine the planet’s overall spectrum, then the pure spectrum of each region must define vertices of a simplex that encloses the data. A simplex is a polyhedron with $N_{\text{dim}} + 1$ vertices. For our two-dimensional PCP, a simplex is a triangle. However, as seen in Figure~\ref{fig:pcp}, the projected spectra hardly represent a triangular structure and would be better enclosed with a quadrilateral. This suggests that we have four end-member spectra.

\section{Comparison with Sonora Diamondback Atmospheric Models}
\label{sec:diamondback}

In this section, we follow a similar approach to \citet{Vos_2023} and \citet{mccarthy2024multiple}, exploring different possible combinations and numbers of models that best fit our NIRISS observations. We utilize the Sonora family Diamondback spectra models --- a new grid of atmospheric models for substellar objects with a range of parameters such as temperature, cloud coverage, and metallicity \citep{Morley_2024}. We use a limited temperature grid range based on our prior knowledge of SIMP\,0136. We also do not constrain gravitational acceleration, because previous studies report values between 100 and 316\,m\,s\(^{-2}\); see Table~\ref{tab:grid_parameters} for the grid parameter values we explored.

Before performing model matching, we compare the integrated flux of Diamondback models to $\sigma T^4$, where $T$ is the reported temperature of a given model spectrum. We find that the large majority of grid spectra have higher integrated temperatures than reported for each model. For our grid range, the average difference is 25\,K, the maximum difference for 900\,K objects is 3\%, which rises to 5\% for 1400\,K objects. We do not find these discrepancies to be an issue, as they are comparable to derived temperature uncertainties from other studies. However, one must be careful when using grid models to infer the target’s temperature.

\begin{table}[htb]
\centering
\caption{Sonora Diamondback Model$^{\rm a}$ Grid Parameters Used to Match NIRISS Spectra of SIMP 0136. }
\label{tab:grid_parameters}
\begin{tabular}{cccc}
\hline \hline
\multicolumn{1}{c}{T (K)} &
    \multicolumn{1}{c}{g (m s$^{-2}$)} &
    \multicolumn{1}{c}{$f_{\rm sed}$} &
    \multicolumn{1}{c}{[M/H]} \\
\hline
900  & 31    & 1   & $-0.5$ \\
1000 & 100   & 2   & 0.0    \\
1100 & 316   & 3   & $+0.5$ \\
1200 & 1000  & 4   & ...     \\
1300 & 3160  & 8   & ...     \\
1400 & ...    & NC  & ...     \\
\hline
\end{tabular}
\\[1.5ex]
{\bf Note.} $f_{\rm sed}$ is the sedimentation efficiency parameter that characterizes cloudiness: smaller number means more clouds. NC refers to clear-atmosphere spectra. All models use [C/O] = 1.0.
\\
$^{\rm a}$ \cite{morley2024sonora}.
\end{table}

Rescaling the observed spectrum to the top of the atmosphere flux requires knowledge of the radius of the object and the distance to it. While distance is precisely known from Gaia Early Data Release 3 (Gaia Collaboration et al. 2023), radius estimations for brown dwarfs are usually inconsistent \citep{adam11}. Atmospheric modeling with BT-Settl \citep{allard13} predict SIMP 0136 to be 0.82\,R$_{\rm Jup}$, while evolutionary model SM08 \citep{saumon08}, given the age estimation, provides 1.22\,R$_{\rm Jup}$ \citep{Gagne2017}. In all our calculations, we adopt the latter, as it gives the most comparable fluxes to Diamondback spectra for the most probable T$_{\text{eff}}$ range. A decrease in radius would greatly increase the emitted flux, thus increasing the spectrum-inferred brightness temperature and T$_{\text{eff}}$ of the best-fit Diamondback model above widely accepted values.

\begin{figure}
    \centering
    \includegraphics[width=0.99\linewidth]{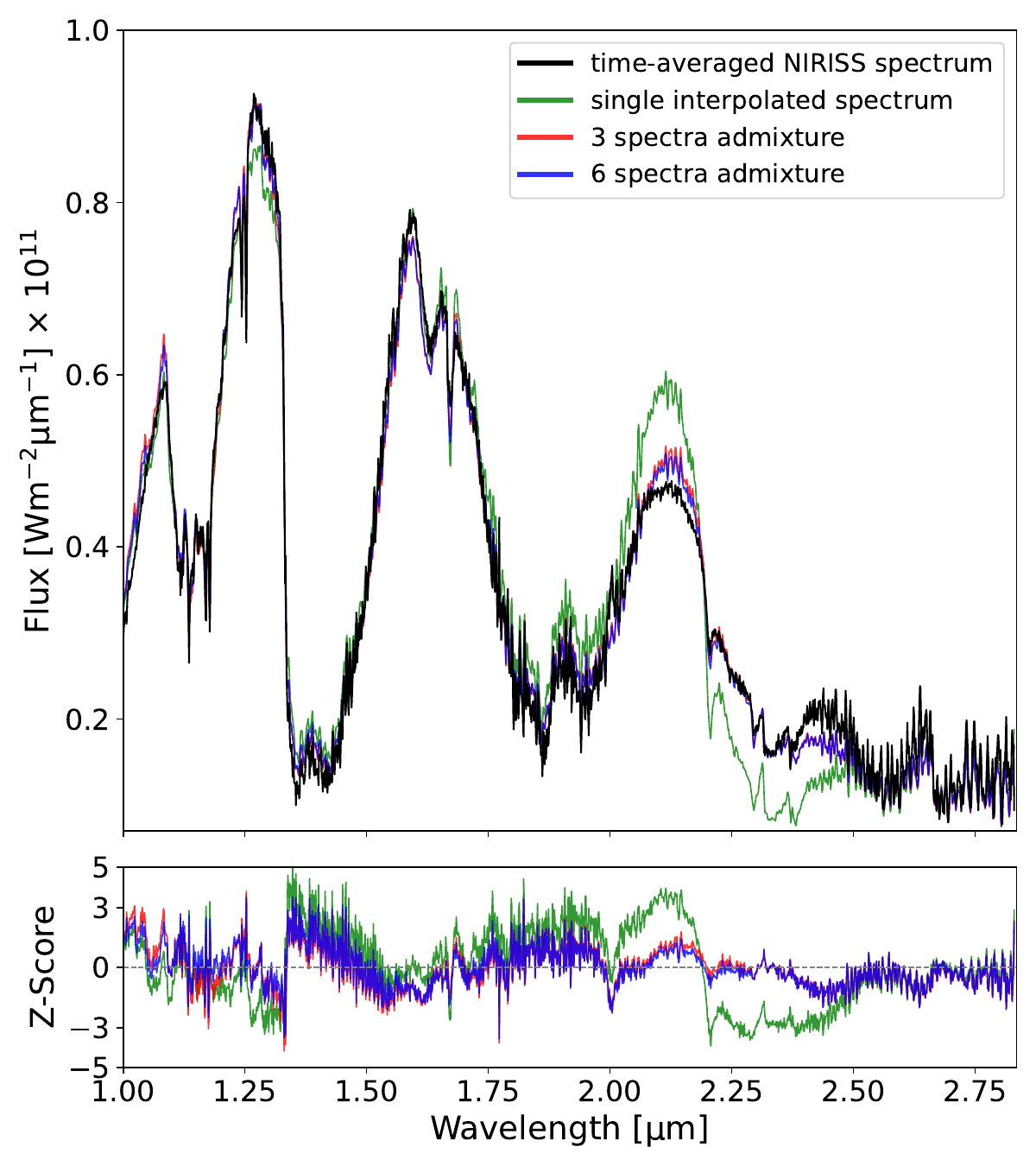}
    \caption{NIRISS time-averaged spectrum over an entire rotation (black line) compared to various Diamondback grid models and their combinations. A single Diamondback model cannot accurately reproduce data. The interpolation of parameters allows the achievement of higher accuracy, with the green line showing a closer match with data. However, only a linear combination of grid models can substantially explain data. The red line shows such a combination of three spectra of different physical properties, and the blue line is an admixture of six models, which is favored by BIC. We allowed gravitational acceleration to vary because the consensus is that it is between 100 and 316 m s\(^{-2}\). The lower part of the figure shows residuals normalized to data error. The error was scaled up by 7.8 to efficiently utilize BIC.}
    \label{fig:1model}
\end{figure}

First, we search for a single model that best fits the time-averaged spectrum. This corresponds to a model with T$_{\text{eff}}=1200$\,K, gravitational acceleration \hbox{g$=100$\,m s\(^{-2}\)}, f$_{\text{sed}}=3$, and metallicity of 0.5. To improve the accuracy, we then create finer-grid interpolations around the best model. The best interpolated spectrum has the following parameters: T$_{\text{eff}} = 1165$\,K, g$=148$\,m\,s\(^{-2}\), f$_{\text{sed}}=3.75$, and metallicity of 0.2. As seen in Figure~\ref{fig:1model}, the interpolated model poorly fits the region around \hbox{2.0--2.5\,$\mu$m}, the 1.25\,$\mu$m and 1.65\,$\mu$m peaks, and the 1.4\,$\mu$m water absorption band. None of the other single grid models described well the time-averaged spectrum or spectrum at any time stamp. This suggests that the observed spectra are linear combinations of multiple distinct spectra.

We used the Bayesian Information Criterion (BIC) to analyze how many models are needed to best describe our data without overfitting. The fitting was performed iteratively: starting with the best-fitting single interpolated model, we added additional models and recalculated weights one at a time to improve the overall fit. The NIRISS spectrum is incredibly precise, and to adequately apply the BIC, we inflated the uncertainties to match the standard deviation of the spectral variability over time. The number of parameters is $N=4 N_{\text{spec}}+(N_{\text{spec}}-1)$, and the BIC is minimized for $N=6$. This could mean that the observed time-averaged spectrum is an admixture of spots that have different temperatures and chemistry. An atmosphere explained by six spectral models is in agreement with previous PCA results, where a majority of the variance is explained by two components, and thus there are $\geq3$ spectral regions that change in time. Combined PCA and BIC analyses are illustrated in Appendix~\ref{sec:pca+bic}. However, without inflating the observational uncertainties, no linear combination of models provides an acceptable match to the data, suggesting that the current models are missing some physical processes and highlighting the need for further refinement.

\begin{figure}
    \centering
    \includegraphics[width=0.99\linewidth]{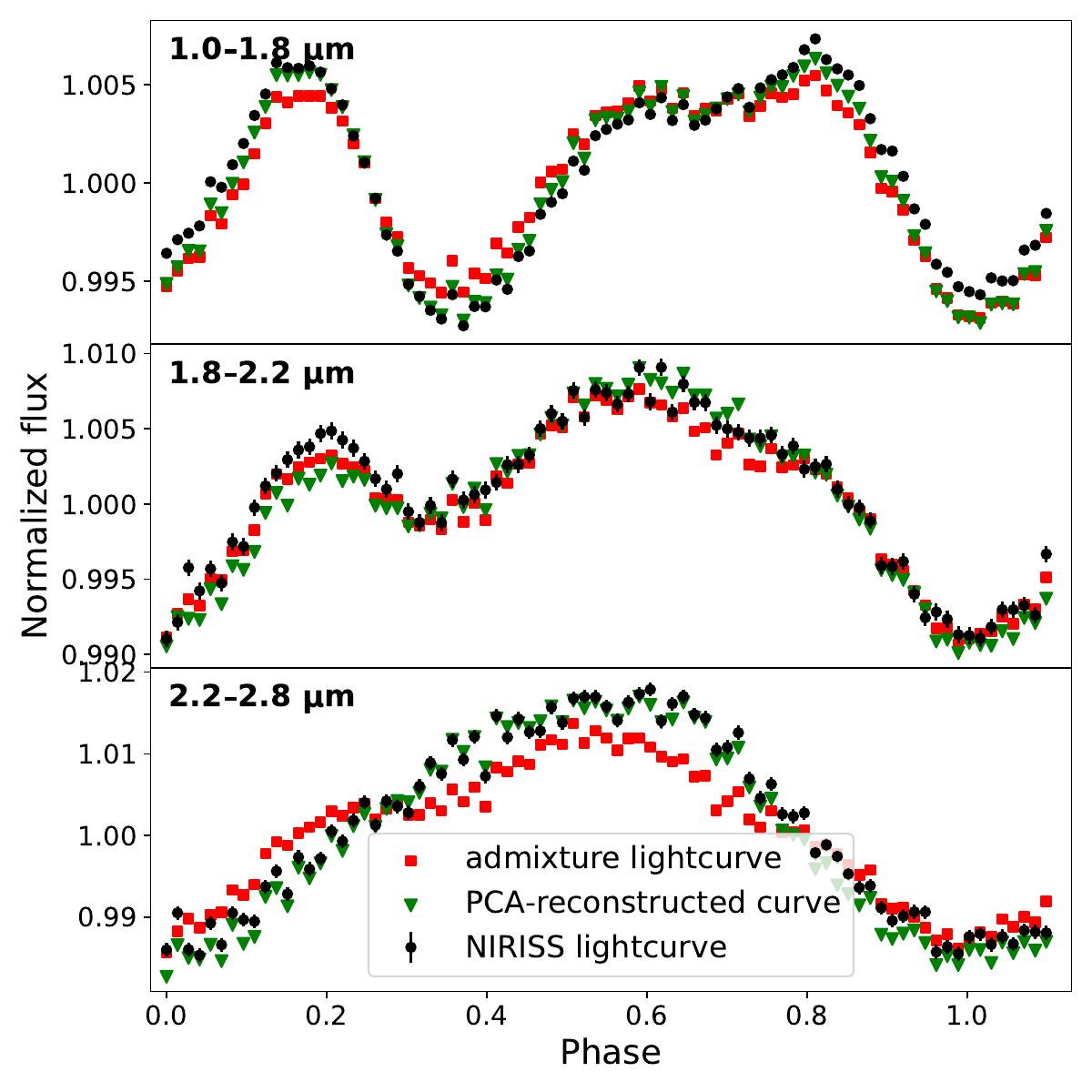}
    \caption{Comparison of lightcurve morphology at three distinct wavelength bins. Black error bars show NIRISS data, similarly to Figure \ref{fig:niriss data}; green triangles show lightcurves extracted from NIRISS data that were projected on the first two principal components. Red squares are lightcurves of an admixture model consisting of three Diamondback spectra. Observed lightcurve morphology can be explained by two components, confirming that it is caused by three or more spectral regions rotating in and out of view. The consistency between the observed and reconstructed lightcurves suggests that the bulk properties of the underlying cloud structure are correctly reproduced.}
    \label{fig:lightcurves_comparison}
\end{figure}

We constrain the number of admixture components to three based on the results of the PCA and find a best-fit combination for matching the time-averaged spectrum. Physical parameters of these three models are: 1000, 1200, and 1300\,K; 3, 2, and 8 f$_{\rm sed}$; 100, 100, and 316\,m\,s\(^{-2}\); and $-0.5$, $+0.5$, and $+0.5$ [M/H]. The integrated Stefan--Boltzmann temperature of the admixture is 1168\,K. As shown by the red curve in Figure~\ref{fig:1model}, this reduced admixture model performs only marginally worse than the more complex six-component fit to the time-averaged spectrum. We then use this three-model combination to fit the time-varying spectra. In Figure~\ref{fig:lightcurves_comparison}, we assess whether the observed spectroscopic variability can be captured by the three-model fit. While it slightly underestimates the flux at certain phases, the overall morphology of the variability is well reproduced.

\section{Atmospheric retrievals and \textit{T-P} profiles}
\label{sec:retrievals}

We performed atmospheric retrievals on the time-averaged spectrum using the Brewster framework \citep{Burningham2017, Burningham2021}. The temperature profile was parameterized following \citet{Madhusudhan2009}. Our model accounts for the opacity contributions of H$_{2}$O, CO, CO$_{2}$, CH$_{4}$, NH$_{3}$, CrH, FeH, Na, and K, which are spectroscopically active in the near-infrared and have been previously identified as key absorbing species in the spectra of L/T transition objects. Additionally, continuum opacities from collision-induced absorption and Rayleigh scattering due to H$_{2}$, He, and CH$_{4}$ are included. Previous retrieval analyses of SIMP 0136 \citep{Vos_2023} indicate the archival 1–15\,$\rm \mu m$ spectral data are best described by a patchy, high-altitude forsterite/enstatite (Mg$_{2}$SiO$_{4}$/MgSiO$_{3}$) cloud overlying a deeper, optically thick iron (Fe) cloud; we follow this prescription.

\begin{figure}
    \centering
    \includegraphics[width=0.99\linewidth]{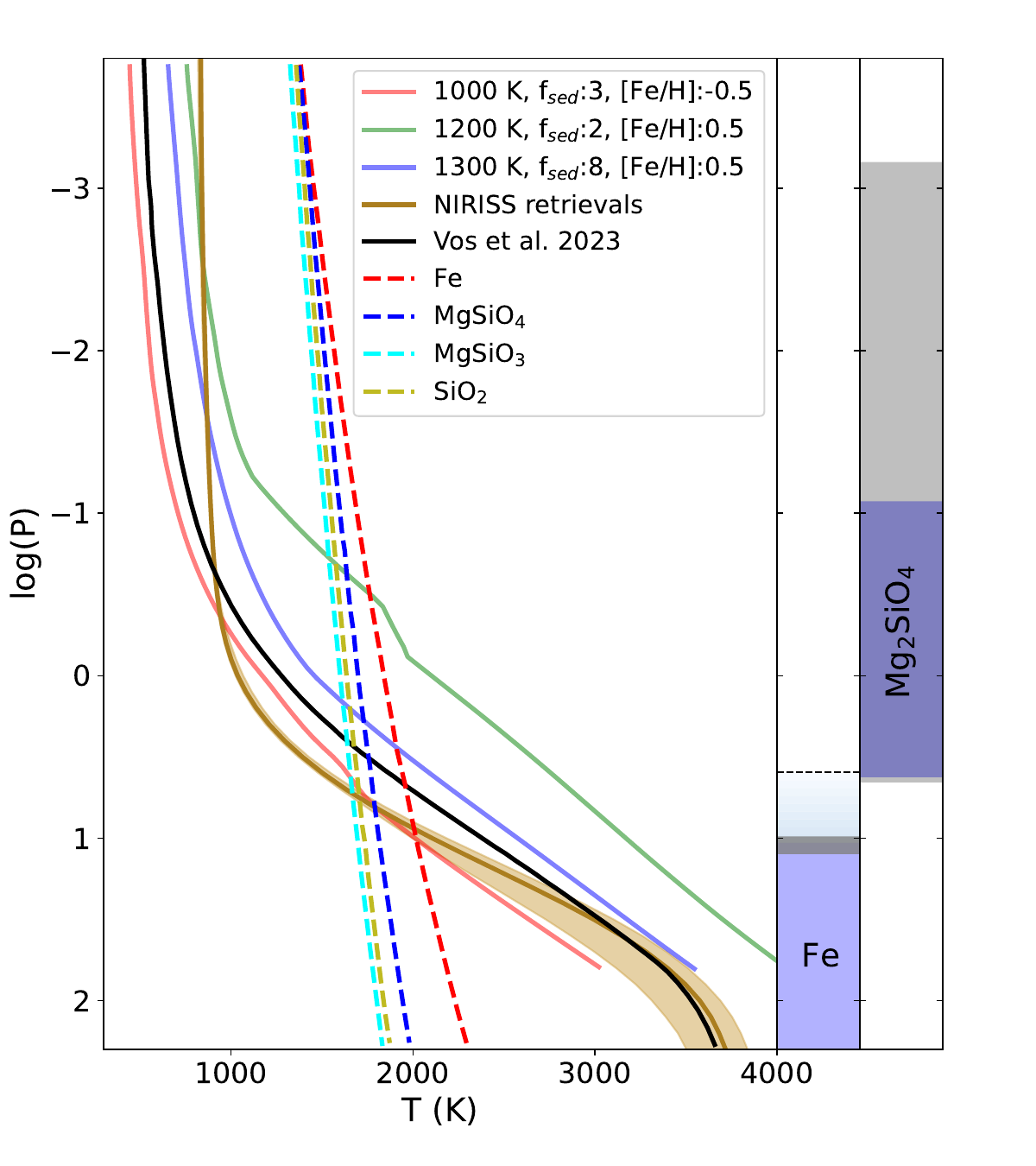}
    \caption{Retrieved Brewster model thermal profile (brown line, with colored shading for 1$\sigma$ confidence interval) for SIMP 0136. Previous retrieval result from \citet{Vos_2023} is plotted as a black curve for comparison. Self-consistent model profiles from the Sonora Diamondback grid are plotted as solid colored lines. Phase-equilibrium condensation curves for plausible cloud species are plotted as colored dashed lines. The clouds pressures are indicated in bars to the right. Purple bar
indicates the median cloud location for the forsterite slab and the optically thick extent of the iron deck, with gray shading indicating the $1\sigma$ range. Graduated shading shows the range
over which the deck cloud optical depth drops to $\rm 1/e$. Forsterite and enstatite are fairly challenging to distinguish in retrievals of the spectra, meaning high-altitude clouds could actually be enstatite or a mixture of both.}
    \label{fig:general_TP}
\end{figure}

Figure~\ref{fig:general_TP} presents the retrieved temperature profile and cloud distributions, compared to Diamondback models (the same used in Figure~\ref{fig:1model}) and phase-equilibrium condensation curves. The retrieved temperature profile (brown curve) shows an adiabatic slope below 1 bar, in good agreement with retrieval results of \cite{Vos_2023} that combine 1--2.5\,$\mu$m Infrared Telescope Facility/SpeX Prism spectrum, a 2.5--5\,$\mu$m AKARI/Infrared Camera spectrum, and a 5--15\,$\mu$m Spitzer/Infrared Spectrograph spectrum. Self-consistent grid models of the best-fit three-spectra admixture model are also plotted for reference with \hbox{T$_{\rm eff}$ = 1000, 1200, 1300\,K}, \hbox{g = 100, 100, 316}, and \hbox{$f_{\rm sed}$ = 3, 2, 8}. While being broadly consistent, our retrieved profile is more isothermal at high altitudes and slightly cooler at 1--10 bars. This discrepancy may be attributed to the limited flux contribution from the upper atmosphere in the near-infrared bands, resulting in weaker constraints on the temperature structure at these altitudes.

\section{Mapping}
\label{sec:mapping}

We may infer some brightness geometry of stars and planets via observed lightcurve variations. This inference is complicated by the fact that the brightness-to-map problem is degenerate: one brightness map results in a single lightcurve, but one lightcurve can be produced by potentially infinite map variants. Knowledge of rotational period, inclination, and limb darkening can help to constrain some parameters.

In this work we use two complementary techniques to map SIMP\,0136: sinusoidal \citep{cowan2008inverting, cowan2013light} and spherical harmonic mapping \citep{cowan2013light, starry}. The first method is algebraically simple and assumes a north--south symmetry of the object. The second uses spherical harmonic basis maps.

\begin{figure*}
    \centering
    \includegraphics[width=1\linewidth]{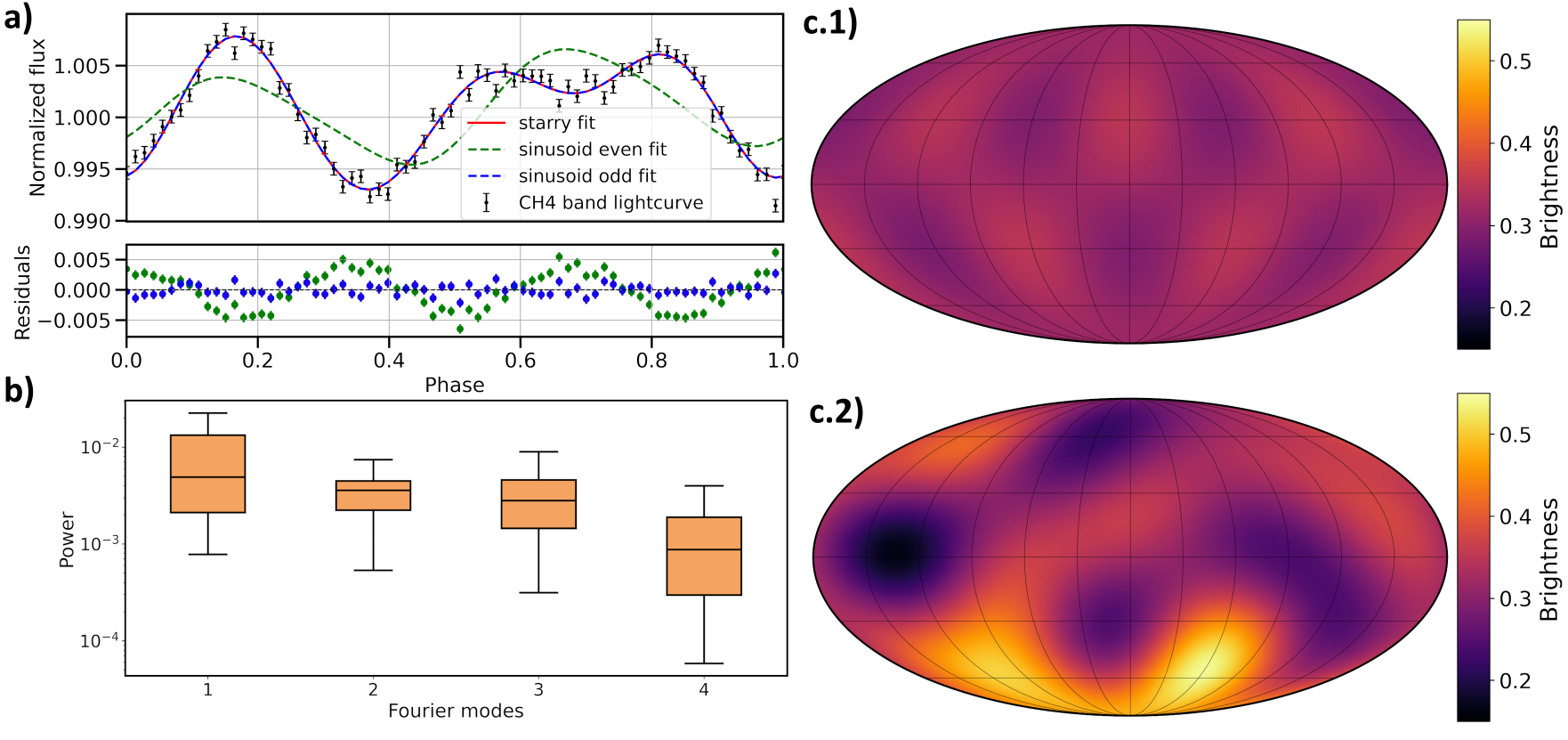}
    \caption{(a) \hbox{1.6 -- 1.8 $\mu m$} band lightcurve (black dots) fitted with different mapping models. Green dashed line: four-mode sinusoidal fit excluding the third mode; blue dashed line: four-mode sinusoids with the odd mode; solid red line: spherical harmonic mapping. Plot below shows residuals of each method; the sinusoidal odd mode fit and spherical harmonics overlap due to identical performance. (b) Power of four Fourier modes from decomposition of all lightcurves. The nonzero third mode confirms the object's inclination and implies there is a north--south asymmetry in the atmosphere of SIMP\,0136. (c.1) Best-fit spherical harmonics map inferred from the $1.0 - 1.31\,\mu$m band lightcurve with the uniform prior on Y$_{\rm lm}$ coefficients. This map demonstrates north--south asymmetry, but is physically implausible. (c.2) A spherical harmonics map of the same lightcurve with Y$_{\rm lm}$ priors of an arbitrary spot position. The first map fits the observed lightcurve with $\chi^2=210$; the second fits with $\chi^2=209.8$. The fact that both maps look different yet produce indistinguishable lightcurves is a testament to the degeneracy of mapping without enough external constrains.} 
    \label{fig:odd_modes+maps}
\end{figure*}

If the thermal map is north--south symmetric, then odd harmonics do not have a phase function signature. Instead, our observations of SIMP 0136 exhibit significant power at the third mode (see Figure~\ref{fig:odd_modes+maps}b), indicating north--south asymmetry and inclined rotation \citep{cowan2013light}.

Spherical harmonics within the \texttt{Starry} framework \citep{starry} allow us to account for the inclination of the object. We provide a multivariate Gaussian prior on the spherical harmonics coefficients with a mean and covariance.

The results from fitting the $1.2$--$1.4\,\mu$m band lightcurve with various mapping methods are shown in Figure~\ref{fig:odd_modes+maps}a. A sinusoidal fit with four modes, excluding the third mode, is represented by the dashed green line. The maximum number of modes was selected using the Bayesian Information Criterion (BIC). This approach failed to capture the distinct shape of the phase curve, nor similar features in other bands. If we include the odd harmonics (dashed blue line), we get a reasonable fit. This means that SIMP 0136 has a north--south asymmetric map.

Spherical harmonic decomposition accounts for inclination and produces 2D maps. Using BIC, we constrained the number of $l$-degrees to four, that is, 25 Y$_{lm}$ harmonics in total. We provided two sets of priors on coefficients: flat with uniform variance, and a spot-band structure. The resulting maps can be seen in Figure~\ref{fig:odd_modes+maps}c.1 and c.2. The asymmetric checkerboard pattern is the most favorable for flat priors. A more plausible map is obtained with a spot-band prior. The fact that these maps produce identical lightcurves is a testament to the extreme mapping degeneracy \citep{cowan2013light}.

\section{Connecting vertical structure to lightcurve morphology}
\label{sec:altitude-var}

SIMP\,0136’s spectrally inhomogeneous variability indicates multiple layers with distinct atmospheric processes and dynamics. The enhanced near-infrared (NIR) spectrophotometric variability observed at the L/T spectral transition has been demonstrated to be best fit by Fourier modes for objects such as WISE\,1049B (aka Luhman\,16B), VHS\,1256\,b, and SIMP 0136\ \citep[e.g.,][]{Apai2017,Apai2021,Fuda2024,Zhou_2020a,Plummer2024b}. In this section, we employ a contribution function (and our previous harmonic analysis above) to map spectral wavelength ranges to pressure levels and create 2D (pressure-versus-time) flux variability maps, which can be decomposed to determine variability sources at each atmospheric layer.

First, we fit the multi-spectral lightcurves with Fourier models using \texttt{Imber} \citep{Plummer&Wang2022,Plummer&Wang2023}, which employs dynamic nested sampling \citep{Skilling2004,Skilling2006} via the open-source Python code \texttt{Dynesty} \citep{Speagle2020}. Considering the spectral band from $1 - 2.8 \ \rm{\mu m}$, we use $0.05 \ \rm{\mu m}$ bins to plot 36 lightcurves (see Figure \ref{fig:niriss data} for an example with 0.2 $\rm{\mu m}$ bins). Qualitatively, 3 characteristic lightcurve shapes can be seen in these results, broadly matching the findings of \citet{Biller2024} and \citet{Chen2025} for WISE\,1049B and \citet{McCarthy2025} for SIMP 0136, which found 3, 4, and 9 characteristic lightcurve shapes, respectively, with JWST/NIRSpec data. For each binned lightcurve, we select the best-fit Fourier model based on the Bayesian Information Criterion.

\par
\begin{figure*}
    \centering
    \includegraphics[width=1\linewidth]{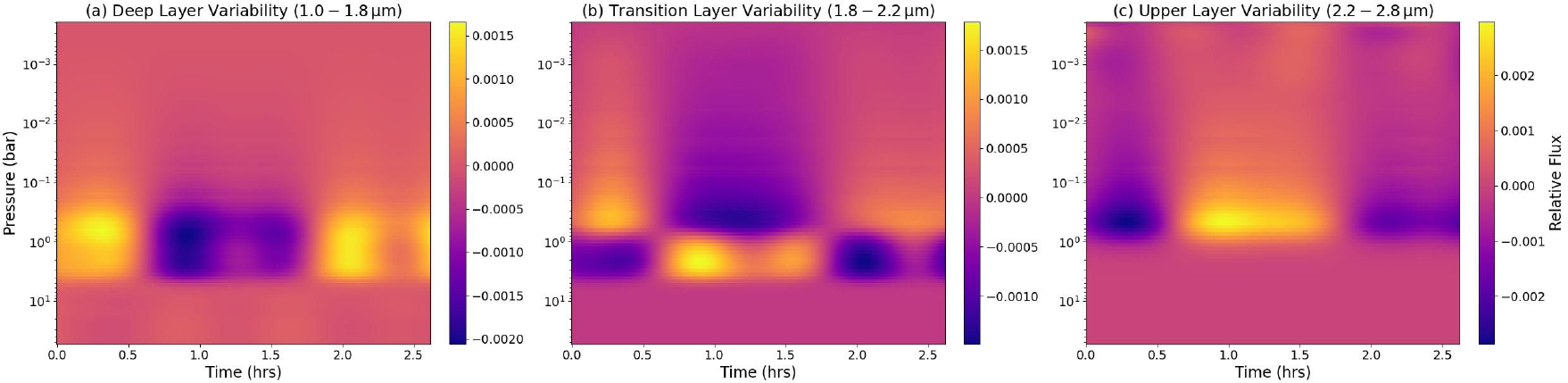}
    \caption{Vertical atmospheric flux (time\slash pressure level) variability maps. ((a)--(c)) Unique variability contribution from each atmospheric layer. (a) Deeper layer containing higher-order harmonics. Variability is interpreted to be due to Mg$_2$SiO$_4$ cloud modulation \citep[e.g.,][]{Vos_2023,McCarthy2024,McCarthy2025,Plummer2024b}. (b) Transition layer containing bright spots colocated with presumed cloud tops in the lower layer. Transition layer corresponds to water absorption bands, and the bright spots may indicate localized heating or variations in water absorption associated with cloud formation (either muting of spectral features or depletion of water). (c) Upper layer demonstrating lower-order harmonics thought to be driven by bright spots in the transition layer. 
    } 
    \label{fig:1D_atmospheric_flux_map}
\end{figure*}

 To create vertical maps, we use the best-fit Fourier models and a 1150\,K, $g = 316~\rm{m~s^{-2}}$ \citep{Gagne2017,Vos_2023}, cloud-free Sonora Bobcat substellar atmospheric model \citep{Marley2021} and \texttt{Picaso} \citep{Picaso}. At each time step, we scale 0.05\,$\mu$m interval spectral bins in the contribution function by the corresponding normalized lightcurves from Figure \ref{fig:niriss data}. We then sum the total emission flux at each pressure level across all wavelengths. This creates a 1D vector of fluxes as a function of pressure. We repeat this process at every time step and then normalize the flux at each pressure level. The resulting vertical flux variability maps (see Figure \ref{fig:1D_atmospheric_flux_map}) suggest three distinct layers, which match the qualitative lightcurve morphology groups seen in Figure \ref{fig:niriss data}.

\par To analyze each layer’s unique harmonic components, we consider the specific contribution from each of its three layers by wavelength ($1.0 - 1.8 \ \rm{\mu m}$, $1.8 - 2.2 \ \rm{\mu m}$, and $2.2 - 2.8 \ \rm{\mu m}$). We do this for each layer by subtracting out the contributions from the other two layers. Each layer’s unique atmospheric flux variability map can be seen in Figure \ref{fig:1D_atmospheric_flux_map}a, b, and c.

\par The deepest layer ($1$--$1.8\,\mu\mathrm{m}$) corresponds to $\gtrsim 1\,\mathrm{bar}$ (Figure~\ref{fig:1D_atmospheric_flux_map}(a)), includes higher-order harmonics ($k=3$, where $k$ is the wavenumber), and corresponds to the spectral region where the models predict cloud variability \citep[e.g.,][]{McCarthy2025}. We interpret higher-order harmonics at this pressure level as forsterite (Mg$_2$SiO$_4$) cloud modulation, in line with \citet{Vos_2023}, \citet{mccarthy2024multiple}, and \citet{Plummer2024b}.

\par The transition layer ($1.8$--$2.2\,\mu\mathrm{m}$) at $\sim 1\,\mathrm{bar}$ corresponds to the expected pressure level of Mg$_2$SiO$_4$ cloud tops \citep[e.g.,][]{Vos_2023, McCarthy2024} and exhibits a transition from waves including complex, higher-order harmonic waves to those dominated by lower-order harmonics. The layer contains bright spots correlated with the suspected cloud features at the top of the deeper layer. This layer corresponds to a prominent water absorption band, and we suspect the bright spot in Figure~\ref{fig:1D_atmospheric_flux_map}(b) depicts either localized heating or variations in observed water absorption tied to the presence of forsterite clouds, either due to muting of spectral features or the depletion of water as part of the thermochemical reaction that produces forsterite clouds \citep{visscher10}.

\par The highest altitude layer ($2.2$--$2.8\,\mu\mathrm{m}$) with pressures $\lesssim 200\,\mathrm{mbar}$ (see Figure~\ref{fig:1D_atmospheric_flux_map}(c)) typically contains lower-order harmonics ($k=1, 2$). This high layer's wavelength range includes both water and carbon monoxide (CO) absorption bands. As can be seen by comparison to Figure~\ref{fig:1D_atmospheric_flux_map}(c), the upper layer's flux variability is in phase with the bright spot in the transition layer at $\sim 1\,\mathrm{bar}$.

\section{Regional spectra retrieval}
\label{sec:reg_spec}

\citet{cubillos2021longitudinally} proposed to invert phase-dependent spectral observations and produce longitudinally resolved spectra that can then be fit using standard 1D spectral retrieval codes. Since {SIMP 0136}'s variability is more accurately described by spherical harmonics maps, we extend the method of regionally resolved spectra to 2D maps. By having time-resolved spectra covering the full rotational phase, it is in principle possible to construct brightness maps at different wavelengths and extract spectra from specific regions on the map.

 In our case, the extracted regional spectra and their corresponding atmospheric retrievals are unreliable due to degeneracies inherent in 2D mapping. We also encountered difficulties due to the absence of grid-based flux in the forward model. As already shown by \citet{cubillos2021longitudinally}, the grid flux normalization is not trivial. Nevertheless, some results can be interpreted. The temperature structure varies significantly across different regions, suggesting that variability is primarily due to temperature inhomogeneity. Some species are well constrained and consistent between regions: water, methane, alkalies, and iron hydride. However, regional spectra show higher water, alkali, and methane abundances than the time-averaged spectrum. At the same time, CO and carbon oxide, ammonia, and chromium hydride have very different and uncertain posterior distributions. This could suggest that carbon chemistry is not the main source of variability. Additional details on the regional spectral retrieval attempts are provided in Appendix~\ref{sec:regional_spec_results}.

Although this approach did not yield robust results, we consider it valuable to document the steps taken and challenges faced, as they may inform future efforts in this area. Notably, for objects exhibiting north--south atmospheric asymmetry, spherical harmonics maps may remain the only viable method to study individual spectral regions—provided that sufficient observational constraints are available.

\section{Doppler Tomography}
\label{sec:doppler}
While the spectral resolution of NIRISS/SOSS is modest compared to ground-based precision radial velocity (pRV) spectrographs, the instrument's stability enables meaningful Doppler measurements. The SOSSISSE framework describes the data as fractional variations relative to a mean flux, presented as a residual map. To measure velocity shifts, we adopt the approach of \citet{Bouchy2001}, in which radial velocity shifts are determined by projecting the residuals onto the derivative of a high-signal-to-noise ratio (SNR) template. Although originally developed for the pRV field, this method is applicable to any spectroscopic time series. The \citet{Bouchy2001} formalism also provides associated uncertainties (see their Equation~(13)).

Given the low spectral resolution of SOSS and its limited constraints on radial velocities, we do not attempt to divide the domain into wavelength bins as is done with flux variability. Instead, we consider the integral of Equation~(9) from \citet{Bouchy2001} across the 1.00--2.35\,$\mu$m range. The redder portion of the spectrum is excluded due to its markedly different lightcurve behavior.

\begin{figure}
    \centering
    \includegraphics[width=\linewidth]{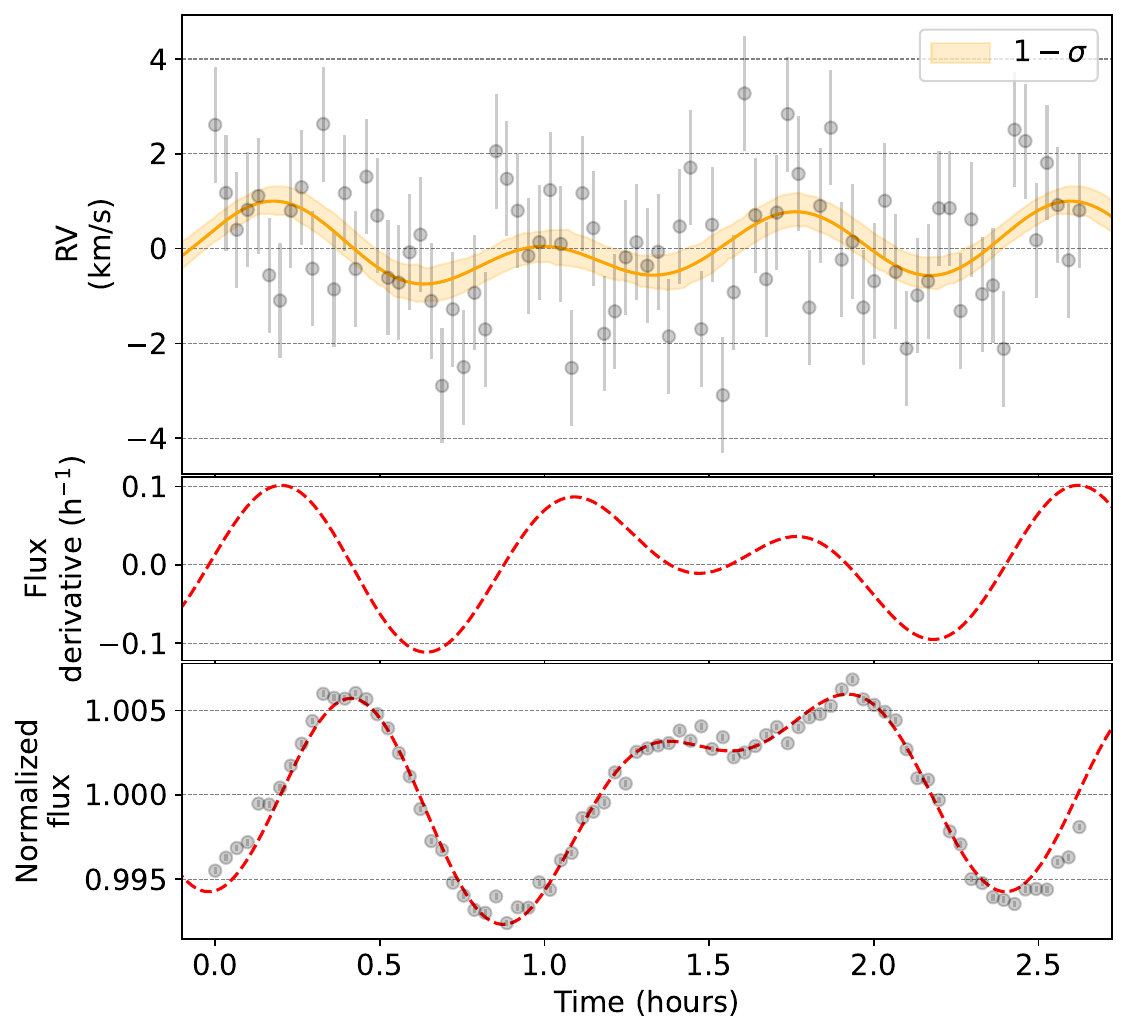}
    \caption{Top: measured and smoothed RV signal at 1--1.35 $\mu$m. Middle: derivative of the lightcurve at the same wavelength bin. Bottom: photometric lightcurve at this bin. The $ff^\prime$ framework suggests that the RV should be in antiphase with the flux derivative. This is the case for the first harmonic only, while second and third harmonics are closely in phase with the flux derivative. }
    \label{fig:rv}
\end{figure}

Figure~\ref{fig:rv} shows the resulting radial velocity time series, revealing correlated variations at the km\,s$^{-1}$ level, characterized by a three-harmonic signal. Per-point uncertainties are approximately 1\,km\,s$^{-1}$. The signal is in phase with the time derivative of the flux, consistent with the framework described by \citet{Aigrain2012}, in which radial velocity (RV) variability for a rotating object (neglecting convective inhibition) scales with the product of the flux and its time derivative---the so-called $ff^\prime$ term in activity-related RV corrections.

The amplitude of RV variations observed for SIMP0136 is consistent with expectations based on its basic physical properties. For a rotating sphere with surface inhomogeneities (e.g., spots), one expects RV jitter on the order of the projected rotational velocity ($v\sin i$) multiplied by the relative photometric variability. With percent-level variability and $v\sin i \sim 50$\,km s$^{-1}$ \citep{Vos2017}, the observed RV jitter matches the expected order of magnitude.

These results underscore an unexplored opportunity in JWST time-series spectroscopy of brown dwarfs. The high precision of molecular-band flux variability measurements---informing cloud structures at different altitudes---can be complemented by RV-based constraints on longitudinal and latitudinal features. While phase-resolved lightcurves constrain longitudinal structures, RV amplitudes are particularly sensitive to low-latitude features, providing strong priors for brightness map reconstructions. Although the SOSS data for SIMP0136 lacks sufficient SNR to allow subdivision by bandpass, higher-resolution modes such as NIRSpec’s G140H or G235H are of particular interest. For a profile limited by instrumental resolution, RV uncertainty scales with the 3/2 power of resolution at a fixed flux level. Thus, moving from $R \sim 700$ (SOSS) to $R \sim 2700$ (G140H/G235H) yields a theoretical sevenfold improvement in SNR, enabling subdivision into several tens of spectral segments while retaining kilometer-per-second-level RV precision.

\section{Discussion and conclusion}
\label{sec:disconcl}
In this work, we presented a multifaceted analysis of time-resolved spectroscopy of the planetary mass object SIMP J01365662+093347. The data were acquired with JWST NIRISS/SOSS and reduced with SOSSISSE. As expected from JWST's stability, and amply demonstrated with transit work, the data are far beyond what could be obtained through ground-based observations of brown dwarfs. This level of detail opens a new window into brown dwarf atmospheric dynamics.

PCA showed that there are at least three different spectral regions rotating in and out of view. Indeed, not a single atmospheric model could fit the observed spectrum, and a mixture of six Diamondback models is needed to adequately fit the time-averaged spectrum. Projected interpolated Diamondback models on the PCP suggest that disk-integrated physical parameters vary in time. These parameters are temperature, metallicity, and sedimentation efficiency. The fact that we have small spectral variations and change in integrated physical parameters in time ($\approx 10$ K, $f_{\rm sed} \pm 0.2$, $[\rm M]/[\rm H] \pm 0.2$), suggests that SIMP 0136 has small-scale cloud patches of different temperature, as predicted by brown dwarf GCM results \citep{Tan2021a}. Discrepancy in the reported temperatures of Diamondback models does not affect the magnitude of temperature variability, as it only creates a linear bias. The inhomogeneity in the integrated effective metallicity may arise from different amounts of chemical disequilibrium in each spectral region, e.g., differences in the CO/CH$_4$ ratios among cloud layers.

Atmospheric retrievals of NIRISS spectra produce \textit{T-P} profiles that deviate from ground-based and Spitzer retrievals in \cite{Vos_2023}. We suspect this is due to a more limited wavelength coverage of our data set. Results of both retrievals confirm an iron cloud deck deeper in the atmosphere nevertheless.

Odd harmonics in the lightcurve of SIMP 0136 and the results from spherical harmonics mapping imply north--south asymmetry, combined with the previously noted inclined rotation. The 2D maps inferred from a single rotation remain degenerate. Nevertheless, lightcurve morphology in combination with Sonora Bobcat contribution functions reveal multidimensional structure of observed variability. We find that the spatial distribution of forsterite clouds below the 1-bar level is anti-correlated with the abundances of water and CO at higher altitudes. This is evidenced by a decrease in disk-integrated emissivity from forsterite layers, accompanied by an increase in emissivity from layers associated with water and CO absorption bands. North--south asymmetry of the brightness map in the $1$--$1.8\,\mu\rm{m}$ domain could directly arise from the spatial distribution of small-scale forsterite cloud patches. It is not guaranteed that this weather pattern is a permanent feature on SIMP 0136: multiperiod back-to-back observations could show how this morphology evolves.

The fact that the meteorology of SIMP 0136 evolves is confirmed by several ground- and space-based observations at different epochs. In \citep{Artigau2009}, $J$-band ($1.25 \pm 0.16\,\mu$m) photometric observations at Observatoire du Mont-Mégantic in 2008 reveal how the power of Fourier modes varies across nights. Phase-folded lightcurves from September 18 and 19 can be well explained by the first two modes, while phase curves of September 16 and 21 require an odd third-mode harmonic. Further $J$-band observations at Lowell Observatory in 2016 \citep{Croll2016} show consistent single-peak lightcurves across four rotations on the night of November 10. On the night of November 12, SIMP 0136 shows almost no variability in $J$-band, indicating that meteorological features have longitudinally homogenized. The next night, however, sawtooth-like variations are consistently observed again, but in anticorrelation with the night of November 10, which suggests a slightly denser distribution of clouds in one hemisphere. \citep{Plummer2024b} analyzed CFHT $J$-band observations obtained in 2012: the lightcurve from October 14 is described by three Fourier modes; however, the following night's observations alternated between Y, J, H, and K bands, limiting the cadence and only allowing two modes to be inferred for each band. A $\approx 90^\circ$ phase shift between H-K and the J-H and H-K color indices was found, which might have indicated different cloud layers. Lastly, we compare our observations with JWST GO \#3548 NIRSpec/BOTS time-series spectroscopy analyzed by \citep{McCarthy2025}. The binned lightcurve in the 0.9--2.1\,$\mu$m range is well explained by three Fourier modes, with the third mode having equal power as the second mode, just like in our observations. Yet, 37 hours after our observations, the first mode has evolved and increased its contribution, leading to a different lightcurve morphology between the 2 visits, which is not explained by a simple shift in phase.

We explored a method of extracting and retrieving spectra using spherical harmonics mapping at different wavelengths. This method is valuable in inferring chemical abundances and thermal profiles of potential hot and cold spots of the brightness maps. We performed spectral retrievals on six regions of interest, but the results are not consistent. This is due to the degeneracy of 2D mapping. Observations spanning multiple rotations or Doppler tomography could decrease these degeneracies. 

We also use JWST medium resolution spectroscopy for Doppler tomography of SIMP 0136. These measurements could eventually help constrain brightness maps and improve mapping fidelity. The search for a discrepancy between Doppler signals and lightcurve Fourier modes would benefit from higher spectral resolution observations.

\begin{acknowledgments}
This work was supported by NSERC Discovery Grant, Canada Research Chair, McDonald Fellowship, Mitacs Globalink Graduate Fellowship, and FRQNT Science en exil program (DOI: \href{https://doi.org/10.69777/355563}{10.69777/355563}). The authors also thank the Trottier Space Institute and l’Institut de recherche sur les exoplanétes for their financial support and dynamic intellectual environment, as well as the Center for research in astrophysics of Quebec (CRAQ)/AstroQuebec. This research was partially funded by the Canadian Space Agency’s James Webb Space Telescope observer’s program.

R.A., N.C., E.A., and M.P. acknowledge Johanna Vos, Caroline Morley, and Allison McCarthy for their thoughtful input and discussions that aided this research.

B.B. and F.W. acknowledge support from UK Research and Innovation Science and Technology Facilities Council [ST/X001091/1].

M.P. acknowledges support from the United States Air Force Academy, Department of Physics and Meteorology. Approved for unlimited public release (United States Air Force, Public Affairs \# USAFA-DF-2025-479). The views expressed in this article are those of the authors and do not necessarily reflect the official policy or position of the United States Air Force Academy, the Air Force, the Department of Defense, or the U.S. Government.

The JWST data presented in this article were obtained from the Mikulski Archive for Space Telescopes at the Space Telescope Science Institute. The specific observation analyzed can be accessed via doi:\href{https://doi.org/10.17909/zryr-vs58}{10.17909/zryr-vs58}.
\end{acknowledgments}

\textit{Facility:} {JWST -  James Webb Space Telescope (NIRISS).}

\bibliography{refs}
\bibliographystyle{aasjournal}

\appendix

\section{PCA explained variance and BIC}
\label{sec:pca+bic}

\begin{figure}[ht]
    \centering
    \includegraphics[width=0.99\linewidth]{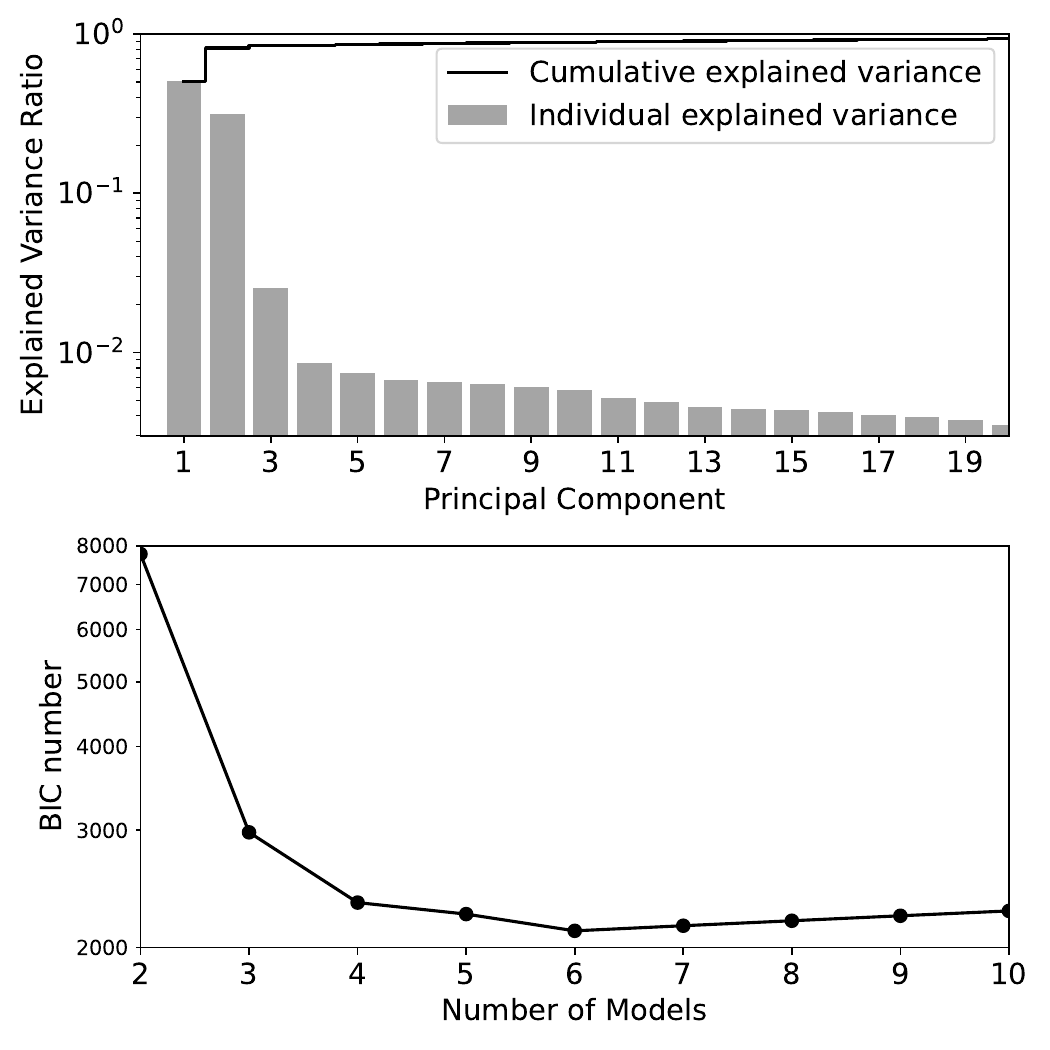}
    \caption{Top: Individual explained variance for each of the principal components shown as a histogram and cumulative explained variance shown with solid  lines. The number of components is limited to 20 for display purposes. Over 80\% of the variance is explained by two principal components, which suggests a minimum of three spectral regions that contribute to spectral variability. Bottom: Analysis of the BIC for admixtures of Diamondback models to fit the time-averaged spectrum. The minimum is reached for a linear combination of six atmospheric models, which is consistent with the PCA.}
    \label{fig:pca_bic}
\end{figure}

Two independent analyses that were used to identify the number of SIMP 0136 variability sources are shown in Figure~\ref{fig:pca_bic}. The first principal component of the PCA explains 50\% of data variance, and the second---31\%, while the rest of the components are insignificant. According to \citet{cowan2011rotational}, this means there are $\geq$3 spectral regions that influence the observed signal variability. This is supported by the BIC analysis during matching linear combinations of Sonora Diamondback models to our time-averaged spectrum, which suggests that six spectral regions is the most optimal case.

\section{Results of the regional spectra retrieval}
\label{sec:regional_spec_results}

\begin{figure}[ht]
    \centering
    \includegraphics[width=0.99\linewidth]{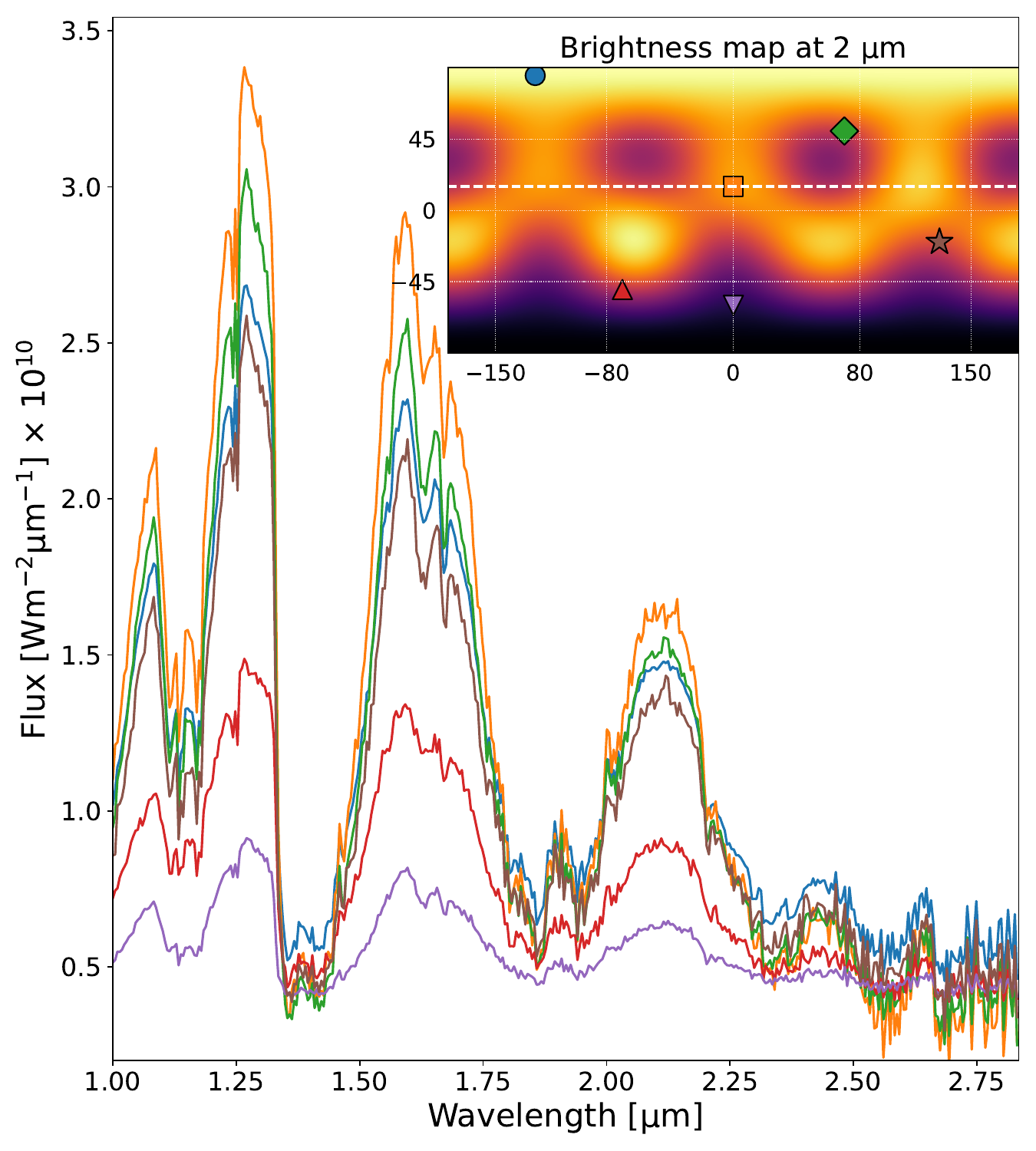}
    \caption{Extracted spectrum from each region of interest is shown on the main plot. The inset map shows the posterior spherical harmonics map at 2 $\mu$m. Each region of interest used in this method is shown with color- and shape-coded data points on the map. The dashed white line indicates subobserver latitude. Although mathematically sounds, the degeneracies involved in 2D mapping make the regional spectra highly uncertain.}
    \label{fig:regional_spectra}
\end{figure}

To avoid uncorrelated noise at individual wavelengths we average the spectra in wavelength bins, then construct maps from lightcurves of binned data. We created maps from 400 lightcurves of 0.0045 $\mu m$ bins using fourth-order spherical harmonics. We then select six regions of interest and extract the spectra from 400 maps at the respective longitude and latitude. The resulting spectra and a map with regions of interest are shown on Fig.\ref{fig:regional_spectra}.

\begin{figure*}
    \centering
    \includegraphics[width=1\linewidth]{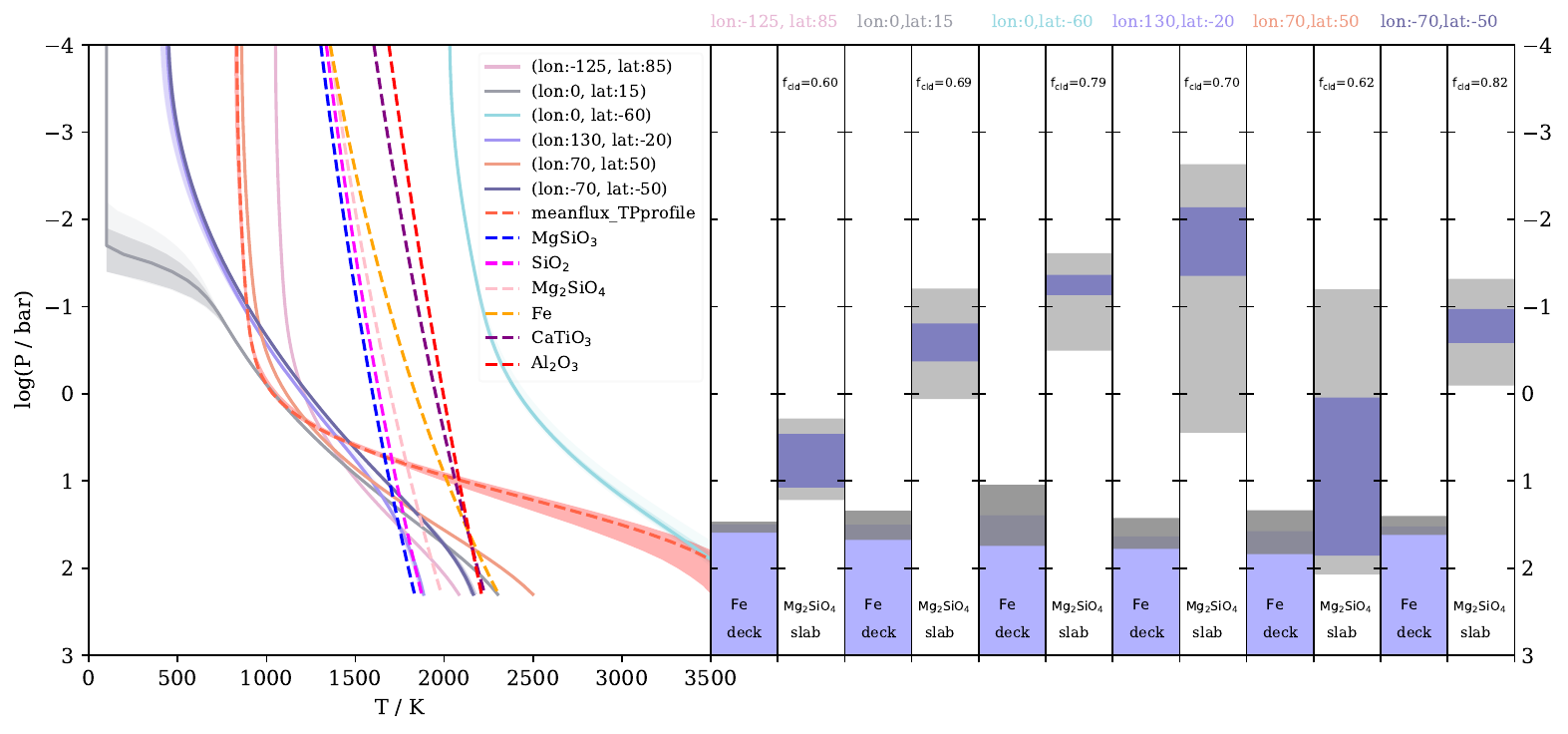}
    \caption{Left: temperature-pressure profiles inferred from atmospheric retrievals of regionally extracted spectra. Red dashed line with 1$\sigma$ confidence interval shows retrieved profile of a time-averaged spectrum. Phase-equilibrium condensation curves for plausible cloud species are plotted as colored
dashed lines. Right: altitude and fraction of patchy forsterite cloud slab and opaque iron cloud deck inferred from \textit{T-P} profiles for each region of interest. Purple bar indicates the median cloud location with gray shading indicating the 1$\sigma$ range.}
    \label{fig:TP_region}
\end{figure*}

\begin{figure*}
    \centering
    \includegraphics[width=1\linewidth]{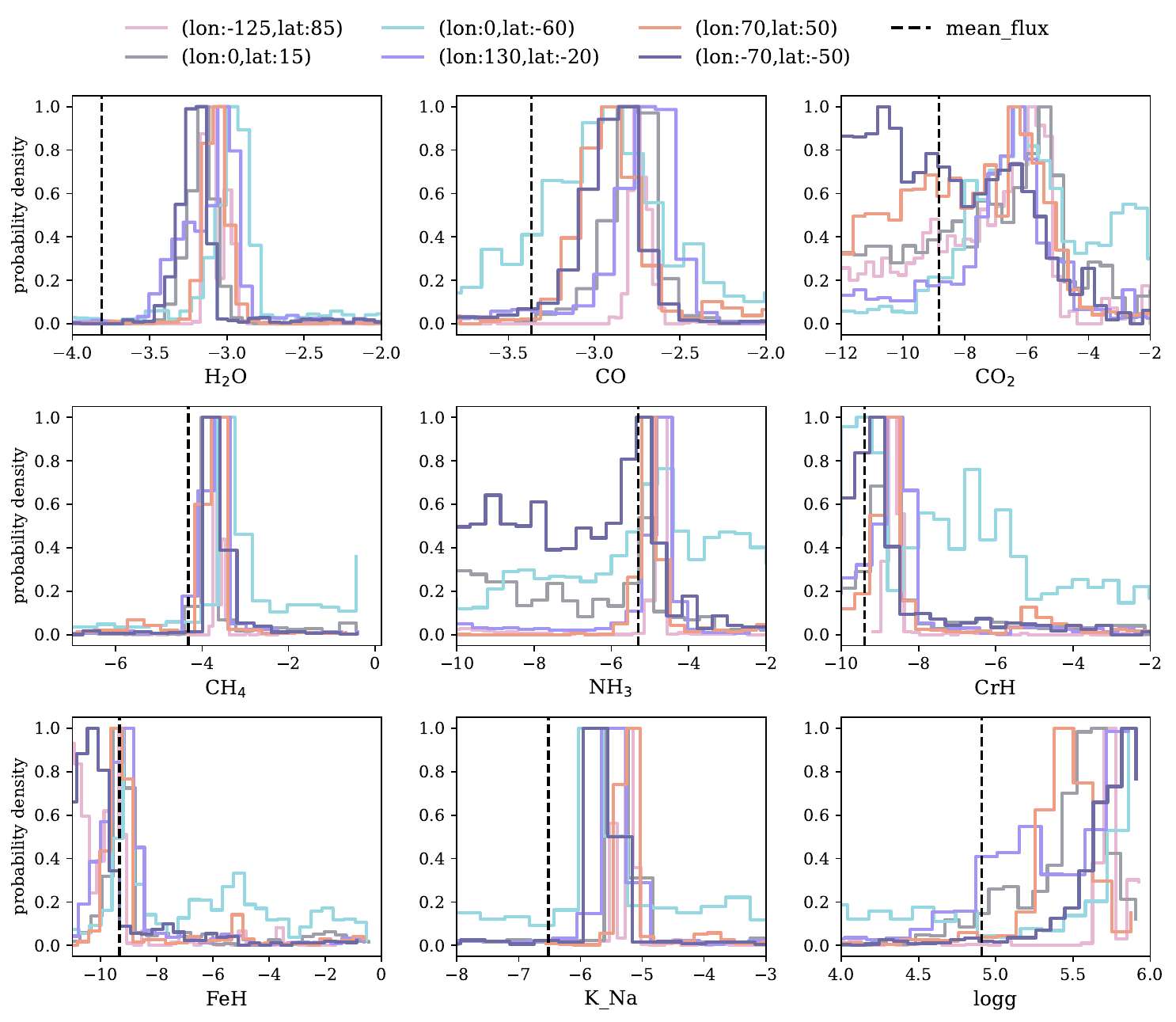}
    \caption{Gas abundances and gravitational acceleration inferred from atmospheric retrieval of regionally extracted spectra. Each posterior histogram is colored according to position on the object. The dashed line is the mean abundances retrieved from the time-averaged spectrum.}
    \label{fig:mol_region}
\end{figure*}

We then apply the same atmospheric retrieval framework as described in Section \ref{sec:retrievals} to explore \textit{T-P} profiles, molecular abundances and cloud levels of each region.  As seen from Fig.\ref{fig:TP_region} the temperature profiles retrieved from regional points tend to be more isothermal in the deep atmosphere, and fail to reproduce the adiabatic slope seen in the \textit{T-P} profile from the time-averaged integrated spectrum retrieval in Section \ref{sec:retrievals}. As a result, the altitudes and fractions of forsterite clouds vary significantly for each region as well. We also detect an anti-correlation, where the darkest cloud patch at latitude: 0, longitude: -60 has the hottest \textit{T-P} profile (cyan curve in Fig.\ref{fig:TP_region}), while the hottest point at latitude: 0, longitude: 15 has the lowest \textit{T-P} profile (grey curve). Molecular abundances are also very different for spectrum of each region (see Figure \ref{fig:mol_region}).

\end{document}